\documentclass[12pt]{iopart}

\usepackage{graphicx}
\usepackage{dcolumn}
\usepackage{bm}
\usepackage{color}

\begin{document}

\title{Activation mechanisms in sodium-doped Silicon MOSFETs}

\author{Ferrus~T, George~R, Barnes~C~H~W, Lumpkin~N, Paul~D~J and Pepper~M}

\address {Cavendish Laboratory, University of Cambridge, Madingley Road, CB3 0HE, Cambridge, United Kingdom}

\ead{taf25@cam.ac.uk}

\begin{abstract}

We have studied the temperature dependence of the conductivity of a silicon MOSFET containing sodium ions in the oxide above 20\,K. We find the impurity band resulting from the presence of charges at the silicon-oxide interface is split into a lower and an upper band. We have observed activation of electrons from the upper band to the conduction band edge as well as from the lower to the upper band. A possible explanation implying the presence of Hubbard bands is given. 

\end{abstract}


\submitto{\JPCM}
                        
\maketitle
  
\section{Introduction}

Contamination of silicon oxides by impurities has represented a long-term problem for Metal-Oxide-Semiconductors (MOS) production as this deeply affects the electrical behaviour and degrades the performance of transistors. Impurities are easily incorporated into silicon or the oxide throughout the fabrication of devices. The first source of contamination occurs during the growth of materials where the temperature facilitates the incorporation of impurities such as sodium, antimony, gallium and phosphorous into oxides and heavy metal like copper, lithium, gold into silicon. However, contamination also happens on processed devices that are left unprotected even in a cleanroom environment
\cite{cleanroom}. The silicon oxide is effectively vulnerable to various fast diffusers at room temperature like H$^+$, OH$^-$ and more specifically Na$^+$ ions. The study of the electronic properties of silicon transistors with oxides containing such types of ions is consequently of great interest from the perspective of optimising the quality of electronic components. Early studies started in the 70's when Fowler and Hartstein used a silicon MOS Field Effect Transistor (MOSFET) to probe the impurity states located below the conduction band in a sodium-contaminated device \cite {Fowler1,Fowler2}. They found the presence of the ions near the Si-SiO$_2$ interface at a concentration of few $10^{12}\,$ions.cm$^{-2}$ was responsible for the formation of an impurity band. At lower impurity concentrations, the situation is more complex as the impurity band splits into a ground and several excited bands \cite{Gold, Erginsoy}.
The formation of two separate bands in sodium-doped MOSFETs was experimentally observed in a previous study for temperatures below 20\,K and confirmed by the analysis of the density of states at different gate voltages (figure 1) \cite{Ferrus}. In the case of sodium, the ground state band is predicted to be formed by a single bound electron while the first excited band is formed a pair of bound electrons (respectively lower and upper Hubbard band). The possibility that two electrons may occupy the same site has been suggested by Fowler et al \cite{UHB} but has remained an open question. This eventuality is of importance to quantum measurement based on charge quantum bits (qubits) \cite {Kane}. In order to investigate that possiblity, it is necessary to understand the electronic properties of low-doped MOS. To this end, we performed complemetary measurements which looked into details of the temperature dependence of the source-drain conductivity in the regime where electrons are activated from the impurity band to the conduction band edge, i.e above 20\,K. We show that results are also consistent with an activation from the lower to the upper band as this has already been observed \cite{Fritzsche2}. Arguments are given in favor of the presence of Hubbard bands.

\section{Experiment}

We have fabricated MOSFETs in a circular geometry (Corbino) from a (100) p-silicon wafer with high resistivity (10$^4$\,$\Omega$.cm). Such transistors has been widely used because of the ability to continuously vary the electron density and the Fermi energy by use of a metal gate. Its geometry  eliminates leakage current paths around the contacts as well as minimizes scattering with Boron acceptor impurities, especially close to the interface. The effective channel length and the interior diameter of the Corbino MOSFETs were measured to be respectively 1 and 110\,$\mu$m. A 35\,nm-gate oxide was grown at 950\,$^{\circ}$C in a dry, chlorine-free oxygen atmosphere. Contacts were realized by implanting phosphorous at high dose and sputtering aluminium. The contact resistivity was measured to be 3.5 and 2.3\,$\Omega$.cm$^{-1}$ respectively at nitrogen and helium temperatures and the sheet resistance was 6.3 and 5.9\,$\Omega$.$\opensquare$ $^{-1}$ for the same temperatures. Sodium ions were introduced onto the oxide surface by immersing the device in a $10^{-7}$\,N solution of high purity sodium chloride in deionized water. The surface of the chip was then dried with nitrogen gas and an aluminium gate subsequently evaporated. The application of a positive gate voltage (+4\,V at $65^\circ$C for 10\,mins) causes the sodium ions to drift towards the Si-SiO$_2$ interface without diffusing into silicon \cite{Snow,Yon}. The application of $-4$\,V dc in the same conditions removes the ions from the interface. The ions are frozen at their position once the device temperature becomes lower than 150\,K (figure1). Standard low-noise lock-in techniques with an amplifier of 10$^6$\,V/A were used to measure the source to drain conductivity. An ac excitation of 15\,$\mu$V and a frequency of 11\,Hz were chosen. The dc offset of the amplifier was cut using appropriate RC filters. Finally, the gate voltage was controlled by a high resolution digital to analog converter and the temperature measured by a calibrated germanium thermometer.

Several devices were processed identically and gave results that lead to identical conclusions although we noticed some variations in the relative positions and in the widths of impurity bands as well as in the conductivity values. We also fabricated a number of control devices that were not exposed to sodium contamination and were used for comparison. The subthreshold to saturation current ratio was about $10^3$ at 300\,K and $10^6$ at 0.3\,K. In the case of doped devices this ratio falls down to 10 at 300\,K and $10^3$ at 0.3\,K. The presence of charges close to the Si-SiO$_2$ interface is responsible for the finite conductivity in the accumulation region of the MOSFET. The following results are presented for a specific device that was chosen for its high reproducibility in time as well as for its high signal to noise ratio.

\begin{figure}
\centering
\resizebox{!}{6cm}{\includegraphics{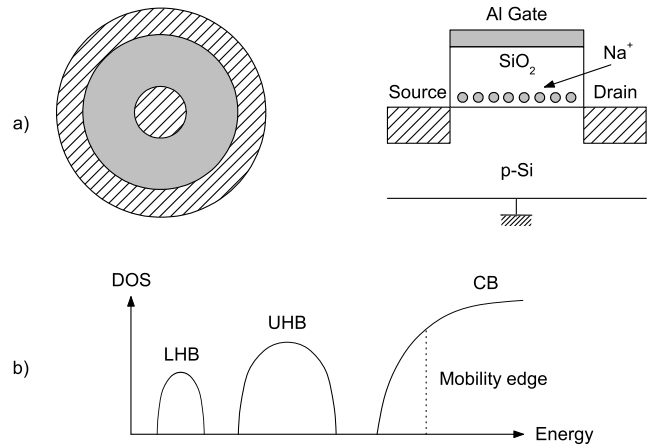}}
\caption{a) Corbino MOSFET used in the experiment when the sodium ions are close to the Si-SiO$_2$ interface. The source is connected at the centre of the Corbino whereas the outside contact is used for the drain b) Schematic diagram of the density of states (DOS) for the present device, with a low energy (LHB) and a high energy (UHB) impurity band separated by a gap to the conduction band edge (CB). The dotted line shows the position of the mobility edge of the conduction band.}
\end{figure}

\section{Results and discussion}

We have measured the source-drain conductivity $\sigma$ for gate voltages $V_g$ between -2.5 and 0.4\,V and for temperatures ranging from $T\,=\,0.3$ to 100\,K. The variation of conductivity with temperature was reproducible 
up to 110\,K. We have previously shown that correlated hopping is present below 20\,K in our device \cite{Ferrus}. In this paper, we focus our analysis on the upper part of the temperature range where hopping is expected to disappear and be replaced by activated behaviour. However, we attempted to fit the curves using a general hopping model in which the exponential prefactor was temperature dependent and the density of state at the Fermi level was energy dependent. Extensive statistical analysis showed hopping does not occur above 20\,K but that the conductivity follows an activated behaviour as described by the expression $\sigma\left(T\right) = \sigma_0\,T^p\,e^{-\epsilon/{k_BT}}$ where $\sigma_0$ and $\epsilon$ are functions of $V_g$ and $p$ is an adjustable parameter. The exponential prefactor was taken to be temperature dependent in order to reflect the temperature dependence of the mobility of electrons \cite{mobility}. Its dependence in temperature as well as the value for $\sigma_0$ reflects the complex scattering mechanisms occuring at the Si-SiO$_2$ interface. Best fits were obtained for $p\,=\,-1$ for which the Arrhenius plot of $\sigma\,T$ gave straight lines, indicating an activation mechanism for the electrons. 

Small-polarons also give rise to such a prefactor in the mobility in the adiabatic regime, where electrons hop without introducing further deformation \cite{Emin, Austin}. However, the highest temperature used in the present study is 100\,K which is well below the energy of longitudinal optical phonons in silicon. The presence of such polarons is thus unlikely. However, this value for $p$ is widely found in systems where the hopping transport is through a random potential or for gaussian density of states \cite{Baranovskii}. This is consistent with early studies on the same device proving the existence of impurities bands \cite{Ferrus}. Such an activated behaviour for the conductivity has already been observed in MOSFETs devices \cite{Fowler1, Mott1}. In our case, the situation is complex and depends on the range of gate voltage studied. For $-0.5$\,V\,$<$\,$V_\textup{g}$\,$<$\,0.2\,V, there exists two activation mechanisms, one at lower temperature characterized by an activation energy $\epsilon_1$ and a second one at higher temperature with an energy $\epsilon_1'$. However, a single activation process with an energy $\epsilon_2$ is found for $-2.5$\,V\,$<$\,$V_\textup{g}$\,$<$\,$-0.5$\,V. For comparison, two impurity bands were found respectively for $-2.5$\,V\,$<$\,$V_\textup{g}$\,$<$\,-1.7\,V (lower band) and for $-1.2$\,V\,$<$\,$V_\textup{g}$\,$<$\,0.2\,V (upper band) (Inset in figure 2).
\begin{figure}
\centering
\resizebox{!}{6cm}{\includegraphics{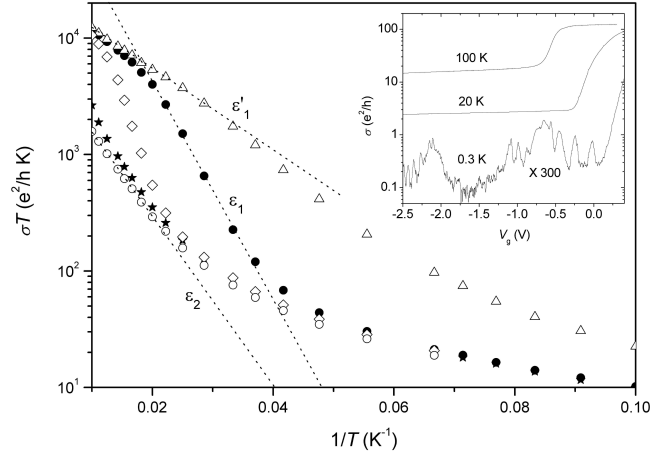}}
\caption{Temperature dependence of the conductivity for $V_\textup{g}$\,=\,0.3\,V ($\opentriangle$),$ -0.15$\,V ($\fullcircle$), $-0.4$\,V ($\opendiamond$), $-0.65$\,V ($\star$) and $-1.95$\,V ($\opencircle$) in the activated regime. The dotted lines defines the different regions of activation and their corresponding energies. The inset shows the variation of log\,$\sigma$ with $V_g$ for T\,=\,100, 20 and 0.3\,K (from top to bottom).}
\end{figure}

Our control devices containing no sodium showed no activated transport at any gate voltage. Below $V_\textup{g}$\,=\,-0.3\,V the conductivity of these devices remained close to zero for the range of temperatures we investigated. Above $V_\textup{g}$\,=\,-0.3\,V and for temperatures up to 50 or 70\,K depending on the gate voltage, only correlated hopping as described by Efros and Shklovskii \cite{Efros} was present (figure 3) and in agreement with Mason and Kravchenko experiments \cite{Mason}. Effectively experiments were carried out in the insulating side of the metal-to-insulator transition where electron-electron interaction may be important in the absence of disorder induced by impurities at the interface. Moreover the size of the Coulomb gap was sufficiently wide not be screened by the temperature. This suggests the presence of activation energies $\epsilon_1'$ and $\epsilon_2$ are related to the presence of impurity traps at the Si-SiO$_2$ interface.

\begin{figure}
\centering
\resizebox{!}{6cm}{\includegraphics{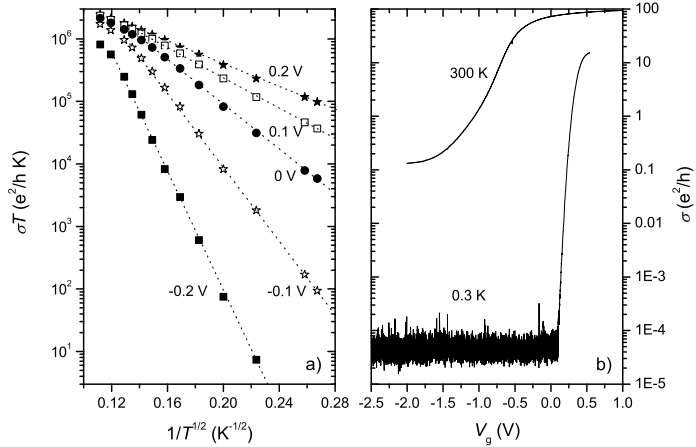}}
\caption{a) Temperature dependence of the conductivity for an undoped device. The difference in threshold voltages between the undoped and the sodium-doped device is 0.2\,V. b) Variation of the conductivity of the undoped device with gate voltage at 300\,K and 0.3\,K.}
\end{figure}

\subsection{Processes $\epsilon_1$ and $\epsilon_1'$}

The presence of two distinct activation energies $\epsilon_1$ and $\epsilon_1'$ suggest two different activation mechanisms. Unlike Fritzsche's observations \cite{Fritzsche} these mechanisms cannot be simultaneous because of the negative curvature of the conductivity in figure 2 at high temperature. Consequently, the source-drain conductivity is not written as a sum of parallel processes in this region. Therefore, this suggests that the mechanism responsible for $\epsilon_1$ may disappear at a given temperature for the benefit of the mechanism $\epsilon_1'$. Figure 4 shows the variation of the activation energy at different voltages. We observe that $\epsilon_1$ has an exponential variation with $V_g$. In MOSFETs, for weak accumulation, the surface potential energy is exponentially dependent in gate voltage. The variation of $\epsilon_1\left(V_g\right)$ then simply reflects the change in the intrinsic Fermi energy and the gate capacitance when the gate voltage is varied. We can then assume that $\epsilon_1$ corresponds to an activation of electrons from the upper band to the conduction band edge \cite{Mott}. The absence of the activation $\epsilon_1$ for $V_g\,<\,-0.55\,V$ may be interpreted as the existence of a conduction band threshold in the upper band separating the localized states in the band tail from a region of conducting states at the centre of the band. The upper band edge is therefore 36\,meV below the conduction band edge. Close to the threshold voltage, the variation of the activation energy with gate voltage is linear, in the first approximation. By capacitance arguments, it is possible to express the value of $\epsilon_1$ in terms of the distance $d_{\textup{Na}^+}$ of the sodium ions from the interface :

\begin{figure}
\centering
\resizebox{!}{6cm}{\includegraphics{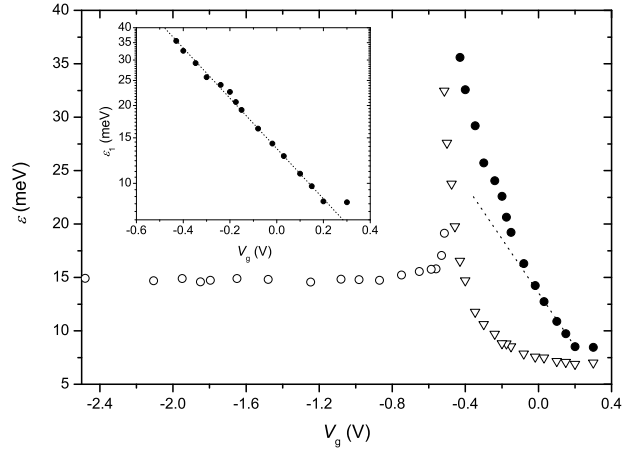}}
\caption{Variation of the activation energy in terms of $V_\textup{g}$ for $\epsilon_1$ ($\fullcircle$), $\epsilon_1'$ ($\opentriangledown$) and $\epsilon_2$ ($\opencircle$). The line represents the linear variation for $V_g$ close to the threshold voltage. Energies are determined within an error $\Delta\epsilon = \pm\,0.5\,$meV for $\epsilon < 25\,$meV and $\Delta\epsilon = \pm\,1\,$meV above. The inset shows the exponential variation of $\epsilon_1$ with gate voltage.}
\end{figure}

\begin{eqnarray}\label{eqn:equation1}
\epsilon_{1} = \epsilon_{1_0}\,-\,e\, \frac{ d_{\textup{Na}^+} } { d_{\textup{ox}} } \left(V_\textup{g}-V_\textup{t}\right)
\end{eqnarray}
\noindent
where $V_\textup{t}$ is the threshold voltage.
\newline\indent
We find that the threshold voltage for conduction in the conduction band is $V_{\textup{t}}\,\sim$\,0.2\,V and that the shallowest localized states of the upper band reside at $\epsilon_{1_0}$\,=\,8.5\,meV below the conduction band edge. The thickness of our oxide being $d_\textup{ox}$ of 35\,nm, the ions may then lie as close as 0.7\,nm from the Si-SiO$_2$ interface. This value is in good agreement with the earlier results from Di Maria \cite{Dimaria} who obtained $d_{\textup{Na}^+}\,=\,0.5\,$nm by measuring the oxide photocurrent at nitrogen temperature. We can also estimate the width of the upper impurity band tail from the energy range for which the activation process is present, giving a width close to 27\,meV. Concerning $\epsilon_1'$, the information obtained from the plot of the activation energy versus gate voltage is not sufficient to determine the nature of the corresponding mechanism and the variation of the pre-exponential factor with $V_g$ (figure 5) needs to be analysed.

\begin{figure}
\centering
\resizebox{!}{6cm}{\includegraphics{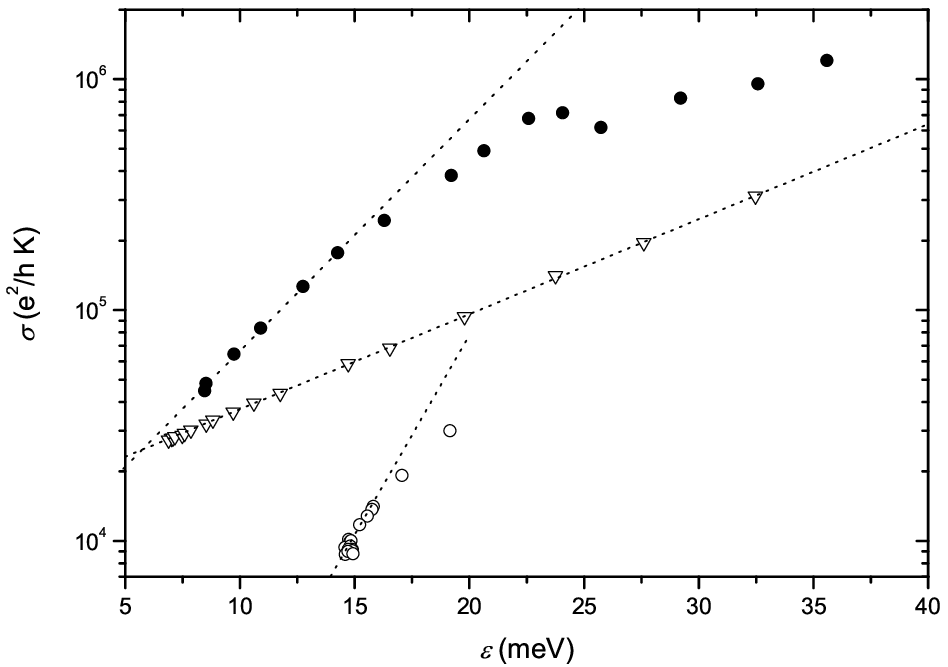}}
\caption{Variation of the exponential prefactor in terms of the activation energy for $\epsilon_1$ ($\fullcircle$), $\epsilon_1'$ ($\opentriangledown$) and $\epsilon_2$ ($\opencircle$).}
\end{figure}

\begin{figure}
\centering
\resizebox{!}{6cm}{\includegraphics{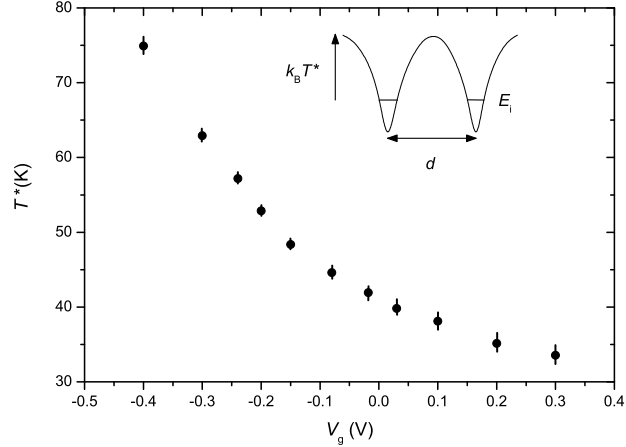}}
\caption{Variation of the critical temperature $T^*$ with $V_g$. Errors are given by the size of bars. The drawing shows two adjacent impurity potentials separated by $d\sim16\pm1\,$nm with $E_{\textup{i}}$ for the energy of the bound state.}
\end{figure}

The plot of $\sigma_{01}'(\epsilon_1')$ (figure 5) is consistent with the Meyer-Neldel rule (MNR) \cite {Meyer} that gives an activation law for the prefactor. This behaviour is found across all ranges of activation energy with a slope corresponding to a typical energy $E_0$ of 10.5\,meV whereas the relation is highly non-monotonic for $\sigma_{01}$. The observation of the MNR is commonly reported in single crystal, polycrystalline, amorphous, organic semiconductors as well as ionic crystals and glasses but more generally in inhomogeneous semiconductors \cite{Roberts, Carlson, Dosdale}. Interpretations on the origin of the MNR are various but the difference in the behaviour of $\sigma_{01}' (\epsilon_1')$ and $\sigma_{01} (\epsilon_1)$ makes our experimental results incompatible with a variation of the Fermi energy in temperature \cite{Popescu}. Multiphonon hopping has also been proposed \cite{Yelon} but this would imply the absence of the MNR for $\epsilon_1'\,<\,E_0$ which is not the case. In an ionic crystal, the MNR is explained by a restructuring of the lattice formed by the ions \cite{Kincs} and an activated conductivity due to an activation of the carrier density or of the mobility \cite{Pepper}. In our case, an analogy could be made except that sodium ions are unlikely to move below 100\,K. We have verified this by performing thermal cycles from 0.3\,K to 100\,K that modify neither the height or the position of the conductivity peaks in figure 2. The mechanism responsible for $\epsilon_1'$ may be interpreted in the following way. Below a certain temperature $T^*$, sodium ions form a disordered lattice and some electrons are localized by the impurity potentials near Si-SiO$_2$ interface at the ion site. The effect of temperature is to thermally activate the bound electrons for conduction with an activation energy $\epsilon_1$ and the plot of $\sigma_{01} (\epsilon_1)$ gives information on the density of states. Above $T^*$, the temperature is sufficiently high to delocalized the electrons from their sites allowing a percolation process to be more efficient and the scattering time to be reduced. Consequently, the conductivity is limited by scattering at the interface and the mobility activated with an activation $\epsilon_1'$. It is then possible to get $\epsilon_1'$ of the order of $k_BT$ or even smaller because the effect of high temperature is nomore to thermally activate carriers but to increase their mobility. This typically describes a process of relaxation happening at the interface like those occurring in melting glasses. As consequence, the slope of $\sigma_{01}'\left(\epsilon_1'\right)$ may give the temperature $T_g$ at which the glass transition occurs \cite{Dyre} and the sodium ions start moving around their positions. We find $T_g\,\sim\,122\,$K. The critical temperature $T^*$ could then correspond to the necessary energy to delocalize the electrons \cite{Martin}. It is experimentally accessible by determining the temperature for which a change in the curvature of $\sigma T$ happens in figure 2. Figure 6 presents the variation of $T^*$ in terms of gate voltage. As expected, the value of $T^*$ decreases when the electron density and the delocalization are increased.

\subsection{Process $\epsilon_2$}

The presence of $\epsilon_2$ is explained by the finite value of the conductivity in the accumulation region of the MOSFET. In undoped devices, the conductivity in the same region was close to zero even at 100\,K. Consequently, we did not find any measurable activation energy in the accumulation region in the reference devices. In the doped device, the average value for $\epsilon_2$ is too small compared to the activation energy $\epsilon_1$ to be related to an activation to the conduction band edge and it is mostly constant over the band gap. It may however results from an activation to an upper state below the conduction band that may be inside the upper impurity band. For $V_g\,<\,-0.5\,$V, the value of $\epsilon_2\,\sim\,15$\,meV is nearly independent of the gate voltage except in the small transition region around -0.55\,V. It is then hardly conceivable that the variation of $\epsilon_2$ is linked to any bandwidth. This behaviour can possibly be explained by considering that the Fermi energy for $V_g$ below -0.55\,V is pinned due to the difference in the number of states in the gap and in the band tails \cite{Adler}. In that case, the Fermi energy then varies abruptly as $V_g$ is made more negative and gets pinned in the lower band. Also, the presence of a band to band activation suggests that the lower band is insulating and that the upper band has a conducting region. In the case of sodium ions in the silicon oxide, the lower band states would then be fully occupied and associated with one bound electron per ion whereas the upper band states would be associated with two electrons per ion. This situation is typically found in Hubbard bands \cite{Norton}. It is then possible to consider that the presence of $\epsilon_2$ is due to an electronic transition from the lower Hubbard band to the upper Hubbard band and that the 15\,meV corresponds to the Hubbard gap. To strenghten this assumption, we need to compare this value to the theoretical estimates in the case of our device. According to the theoretical results of Bethe \cite{Bethe} on the D$^-$ state (i.e. two electrons bound to a donor), an activation from the neutral state (sodium ion with a single electron) to an excited Na$^-$ state is possible. Taking 11.7 for the relative permittivity of silicon as well as 0.19\,$m_{e^-}$ for the effective mass of electrons in silicon, we obtained $\epsilon_2\,=\,$17.8\,meV using Bethe's formula for a single D$^-$ level. Because of the formation of a band the previous estimate for the activation is lowered \cite {Nishimura} and we obtain $\epsilon_2\,\sim\,$15\,meV by taking the value of the localization length $\xi\,\sim\,23\,$nm measured previously on the same device \cite {Ferrus} and a density of neutral donor $N\,\sim\,1.3\times10^{11}\,$cm$^{-2}$. This value is reasonable knowing the average density of donor estimated from the threshold voltage shift gives $N_D\,\sim\,3.7\times10^{11}\,$cm$^{-2}$. Using Nishimura's estimate, we also found the upper states in the D$^-$ band are 3.9\,meV below the conduction band. This value has to be compared with the 8.5\,meV found for $\epsilon_1$ at the threshold voltage. The difference typically gives the position in energy of the conduction band edge relative to the bottom of the conduction band. In order to bring further arguments in support of the existence of a D$^-$ band, it is necessary to give an estimate of the value of the on-site Coulomb energy $U$ using the same method as Schiff but in two dimensions \cite{Schiff}:

\begin{eqnarray}\label{eqn:equation2}
U\,=\, \frac{e^2}{4\pi \epsilon_0 \epsilon \xi}
\end{eqnarray}
\noindent
where $\xi$ is the localization length.
\newline\indent
The value of the localization length as well as its variation in gate voltage was calculated for the same device under the same experimental conditions \cite{Ferrus}. Below $V_g\,=\,-0.5\,$V, $\xi\,\simeq\,22.5\,\pm\,0.7\,$nm and the on-site repulsion energy is $U\,\simeq\,5.2\,$ meV. Supposing a linear relation between the gate voltage and the activation energy, it is possible to convert gate voltages into energies in the gap region at a rate of 7.7\,meV/V (figure 4). The energy $U$ then corresponds to a difference of 0.67\,V in the gate voltage. This agrees well with the value of the existing soft gap between $V_g\,=\,-1.9$ and $-1.2\,$V in figure 2. We can also check the validity of our assumptions by calculating independently the value of the Hubbard gap. For sufficiently low electron density, $T^*$ and $T_g$ are expected to have similar values. The Hubbard gap corresponds to the energy necessary to put a second electron on the same site, that is $E_0+U$. We find a value of 15.7\,meV in agreement with the value of $\epsilon_2$. These observations are thus compatible with the formation of Hubbard bands and a Mott-Hubbard gap formed by Coulomb interactions. Finally, we would like to point out that the formation of D- states in our device may result from a complex interplay between inter-site Coulomb interactions, sodium density and disorder. The inter-site interactions are due to Coulomb interaction between trapped electrons in silicon. These electrons are localized in silicon at the potential minima created by the ions in the oxide. Disorder comes from the fluctuations in position of these minima in the plane but also from the height of the potential wells. Recent studies in GaAs/AlGaAs devices with a silicon d-doped layer showed that disorder could be controlled by the position of the dopant layer relatively to the 2D electron gas \cite{Ghosh}. This indicates that the disorder strength in our device is at least partly related to the position of the ions relatively to the interface.

\section{Conclusion}

We have observed three activation mechanisms in sodium-doped silicon MOSFETs for 20\,K\,$<\,T\,<\,$100\,K. These results are consistent with the existence of two Hubbard-like bands, one with one electron per site and the second with two electrons per site \cite{Norton}. The first mechanism is an activation of electrons from the upper band edge to the conduction band edge. The second takes over the first one when the temperature is sufficiently high to delocalize electrons and corresponds to the activation of the source drain mobility. The last mechanism has been attributed to an activation of electrons from a lower band to an upper band. The theoretical expectations for the position of the bands in energy as well as the soft gap in a case of a D$^-$ band well agree with the values obtained experimentally. It is thus likely that the transport in such a localized system could be explained within the Hubbard model and that the observed upper and lower impurity bands could correspond to Hubbard bands.

\section*{Acknowledgement}

We would like to thank Drs T. Bouchet and F. Torregrossa from Ion Beam System-France for the process in the device as well as funding from the U.S. ARDA through U.S. ARO grant number DAAD19-01-1-0552.

\end{document}